\pdfoutput=1

\documentclass[12pt]{iopart}

\usepackage{xspace}
\def\pythonexists{}
\usepackage{vistrails}
\usepackage{color}
\usepackage{hyperref}
\usepackage{multicol}
\usepackage{listings}

\renewcommand{\vistrailspath}{http://alps.comp-phys.org/vistrails/run_vistrails.php}
\renewcommand{\vistrailsdownload}{http://alps.comp-phys.org/vistrails/download.php}

\newcommand{\eg}{e.g.,\xspace}
\newcommand{\ie}{i.e.,\xspace}
\newcommand{\etc}{etc.\@\xspace}

\begin{document}
\lstset{language=Python,basicstyle=\scriptsize,numbers=left,stepnumber=5,numberfirstline=true}

\title{The ALPS project release 2.0: \\ Open source software for strongly correlated systems}

\newcounter{affiliation}
\newcommand{\myauthor}[3]{#2$^{#1}$}
\newcommand{\myaddress}[2]{\address{\refstepcounter{affiliation} $^{\arabic{affiliation}}$#2 \label{#1}}}

\author{
	\myauthor{\ref{eth}}{B. Bauer}{bauerb@phys.ethz.ch}
	\myauthor{\ref{colorado}}{L.~D.~Carr}{lcarr@mines.edu}
	\myauthor{\ref{graz}}{H.G. Evertz}{evertz@tugraz.at}
	\myauthor{\ref{wyoming}}{A. Feiguin}{afeiguin@uwyo.edu}
	\myauthor{\ref{utah}}{J. Freire}{juliana@cs.utah.edu}
	\myauthor{\ref{goettingen}}{S.~Fuchs}{fuchs@theorie.physik.uni-goettingen.de}
	\myauthor{\ref{eth}}{L.~Gamper}{gamperl@gmail.com}
	\myauthor{\ref{eth}}{J. Gukelberger}{gukelberger@phys.ethz.ch}
	\myauthor{\ref{columbia}}{E. Gull}{gull@phys.columbia.edu}
	\myauthor{\ref{bonn}}{S.~Guertler}{guertler@th.physik.uni-bonn.de}
	\myauthor{\ref{eth}}{A.~Hehn}{hehn@phys.ethz.ch}
	\myauthor{\ref{jaea},\ref{crest}}{R.~Igarashi}{rigarash@hosi.phys.s.u-tokyo.ac.jp}
	\myauthor{\ref{eth}}{S.V.~Isakov}{isakov@phys.ethz.ch}
	\myauthor{\ref{utah}}{D. Koop}{dakoop@cs.utah.edu}
	\myauthor{\ref{eth}}{P.N. Ma}{pingnang@phys.ethz.ch}
	\myauthor{\ref{eth},\ref{utah}}{P.~Mates}{mates@sci.utah.edu}
	\myauthor{\ref{tokyo}}{H.~Matsuo}{halm@looper.t.u-tokyo.ac.jp}
	\myauthor{\ref{paris}}{O. Parcollet}{parcolle@spht.saclay.cea.fr}
	\myauthor{\ref{affpol}}{G.~Paw{\l}owski}{gpawlo@amu.edu.pl}
	\myauthor{\ref{epfl}}{J.D.~Picon}{jean-david.picon@epfl.chl}
	\myauthor{\ref{eth},\ref{harvard}}{L.~Pollet}{pollet@phys.ethz.ch}
	\myauthor{\ref{utah}}{E.~Santos}{emanuele@sci.utah.edu}
	\myauthor{\ref{virginia}}{V.W.~Scarola}{scarola@vt.edu}
	\myauthor{\ref{lmu}}{U.~Schollw\"ock}{schollwoeck@lmu.de}
	\myauthor{\ref{utah}}{C.~Silva}{csilva@sci.utah.edu}
	\myauthor{\ref{eth}}{B.~Surer}{surerb@phys.ethz.ch}
	\myauthor{\ref{crest},\ref{tokyo}}{S. Todo}{wistaria@ap.t.u-tokyo.ac.jp}
	\myauthor{\ref{stationq}}{S.~Trebst}{trebst@kitp.ucsb.edu}
	\myauthor{\ref{eth}}{M.~Troyer}{troyer@ethz.ch}\footnote{Corresponding author: troyer@comp-phys.org}
	\myauthor{\ref{colorado}}{M.~L.~Wall}{mwall@mymail.mines.edu}
	\myauthor{\ref{eth}}{P.~Werner}{werner@phys.ethz.ch}
	\myauthor{\ref{rwth},\ref{stuttgart}}{S. Wessel}{wessel@phys.ethz.ch}
}

\myaddress{eth}{Theoretische Physik, ETH Zurich, 8093 Zurich, Switzerland}
\myaddress{colorado}{Department of Physics, Colorado School of Mines, Golden, CO 80401, USA}
\myaddress{graz}{Institut f\"ur Theoretische Physik, Technische Universit\"at Graz, A-8010 Graz, Austria}
\myaddress{wyoming}{Department of Physics and Astronomy, University of Wyoming, Laramie, Wyoming 82071, USA}
\myaddress{utah}{Scientific Computing and Imaging Institute, University of Utah, Salt Lake City, Utah 84112, USA}
\myaddress{goettingen}{Institut f\"ur Theoretische Physik, Georg-August-Universit\"{a}t G\"{o}ttingen, G\"{o}ttingen, Germany}
\myaddress{columbia}{Columbia University, New York, NY 10027, USA}
\myaddress{bonn}{Bethe Center for Theoretical Physics, Universit\"{a}t Bonn,
 Nussallee 12, 53115 Bonn, Germany}
\myaddress{jaea}{Center for Computational Science \& e-Systems, Japan Atomic Energy Agency, 110-0015 Tokyo, Japan}
\myaddress{crest}{Core Research for Evolutional Science and Technology, Japan Science and Technology Agency, 332-0012 Kawaguchi, Japan}
\myaddress{tokyo}{Department of Applied Physics, University of Tokyo, 113-8656 Tokyo, Japan}
\myaddress{paris}{Institut de Physique Th\'eorique, CEA/DSM/IPhT-CNRS/URA 2306, CEA-Saclay, F-91191 Gif-sur-Yvette, France} 
\myaddress{affpol}{Faculty of Physics, A. Mickiewicz University, Umultowska 85, 61-614 Pozna\'{n}, Poland}
\myaddress{epfl}{Institute of Theoretical Physics, EPF Lausanne, CH-1015 Lausanne, Switzerland}
\myaddress{harvard}{Physics Department, Harvard University, Cambridge 02138, Massachusetts, USA} 
\myaddress{virginia}{Department of Physics, Virginia Tech, Blacksburg, Virginia 24061, USA}
\myaddress{lmu}{Department for Physics, Arnold Sommerfeld Center for Theoretical Physics and Center for NanoScience, University of Munich, 80333 Munich, Germany}
\myaddress{stationq}{Microsoft Research, Station Q, University of California, Santa Barbara, CA 93106, USA}
\myaddress{rwth}{Institute for Solid State Theory, RWTH Aachen University, 52056 Aachen, Germany}
\myaddress{stuttgart}{Institut f\"ur Theoretische Physik III, Universit\"at Stuttgart, Pfaffenwaldring 57, 70550 Stuttgart, Germany}

\begin{abstract}
We present release 2.0 of the ALPS (Algorithms and Libraries for Physics Simulations)
project, an open source software project to develop
libraries and application programs for the simulation of strongly
correlated quantum lattice models such as quantum magnets, lattice
bosons, and strongly correlated fermion systems. The code development is
centered on common XML and HDF5 data formats, libraries to
simplify and speed up code development, common evaluation and plotting tools, and
simulation programs. The programs enable non-experts to start carrying
out serial or parallel numerical simulations by providing basic implementations of the
important algorithms for quantum lattice models: classical and quantum
Monte Carlo (QMC) using non-local updates, extended ensemble
simulations, exact and full diagonalization (ED), the
density matrix renormalization group (DMRG) both in a static version and a dynamic time-evolving block decimation (TEBD) code, and quantum Monte Carlo solvers for dynamical mean field theory (DMFT). 
The ALPS libraries provide a powerful framework for programers to develop their own applications, 
which, for instance, greatly simplify the steps of porting a serial code onto a parallel, distributed memory machine. 
Major changes in release 2.0 include the use of HDF5 for binary data, evaluation tools in Python, support for the Windows operating system, the use of CMake as build system and binary installation packages for Mac OS X and Windows, and integration with the VisTrails workflow provenance tool.
The software is available from our web server at \url{http://alps.comp-phys.org/}.
\end{abstract}


\section{Introduction}
\label{}

In this paper we present release 2.0 of the ALPS project (Algorithms and Libraries for Physics Simulations), an open source software development project for strongly correlated lattice models. This paper updates the publications on the previous releases~\cite{ALPS1.2,ALPS1.3} to include changes and new features in version 2.0.

Quantum fluctuations and competing interactions in quantum many body
systems lead to unusual and exciting properties of strongly correlated
materials such as quantum magnetism, high temperature
superconductivity, heavy fermion
behavior, and topological quantum order.
The strong interactions make accurate analytical treatments hard and 
direct numerical simulations are essential to increase our understanding of the unusual
properties of these systems. 

The last two decades have seen tremendous progress in the development of
algorithms, including refinements to existing numerical algorithms, such as exact diagonalization \cite{lanczos},
the density matrix renormalization group \cite{White1992,Schollwock2005,Daley2004,White2004} and (quantum) 
Monte Carlo (QMC) techniques, as well as the advent of a new class of variational approaches based on 
tensor network states \cite{vidal1,vidal2,Verstraete04,Murg07,vidal07}. 
Some of the algorithmic refinements have led to computational speedups of many orders of magnitude, such as 
the development of non-local update schemes \cite{Swendsen87,Wolff89} in QMC methods \cite{Evertz93,Prokofev98A,Sandvik99,Todo01,looper,Sylyuasen, Evertz03,Alet2005}, 
the sampling of extended ensembles \cite{Wang01,Wang01b,Troyer03,Trebst04,Katzgraber06,Wessel07},
or the continuous-time approach to quantum impurity problems \cite{Rubtsov04,Rubtsov05,Werner06,Werner06Kondo, Gull08_ctaux}.
These advances often come at the cost of increased algorithmic
complexity and challenge the current model of program development in
this research field. In contrast to other research areas, in which
large ``community codes'' are being used, the field of strongly
correlated systems has so far been based mostly on single codes developed by
individual researchers for particular projects. While simple
algorithms used a decade ago could be easily programmed by a beginning
graduate student in a matter of weeks, it now takes substantially
longer to master and implement the new algorithms. At the same time, the use of numerical approaches is increasing. 

\section{The ALPS project}

\subsection{Overview}
The ALPS project aims to
overcome the problems posed by the growing complexity of algorithms
and the specialization of researchers to single algorithms through
an open-source software development initiative. Its goals are to simplify the development of new codes through libraries and evaluation tools, and to help users with ``black box'' codes of some of the most popular algorithms. To achieve these goals the ALPS project provides:
\begin{itemize}
\item {\bf Standardized file formats} based on XML~\cite{xml} and HDF5~\cite{hdf5} to simplify exchange,
distribution and archiving of simulation results, and to achieve
interoperability between codes.

\item {\bf Evaluation tools} for reading, writing and post-processing simulation results, including the creation of 2D and 3D plots.

\item {\bf Libraries} for common aspects of
simulations of quantum and classical lattice models, to simplify code
development and allow implementation on a variety of serial and parallel platforms.
\item A set of {\bf applications} covering the major algorithms. These are useful for: 
{\it theoreticians} who want to test theoretical ideas or new algorithms for
lattice models and to explore their properties, 
{\it experimentalists} trying to fit experimental data to theoretical
models to obtain information about the microscopic properties of
materials, and {\it students} learning computational physics and many-body theory.
\item A {\bf build system} allowing ALPS to be built on Linux, MacOS X, Windows, and supercomputing platforms, and the creation of simple binary installers.
\item An extensive set of {\bf tutorials} teaching the use of the ALPS libraries and the various applications.

\item{\bf License} conditions~\cite{librarylicense,applicationlicense} that encourage researchers to contribute
to the ALPS project by gaining scientific credit for use of their
work.
\item {\bf Outreach} through a web page~\cite{alps}, mailing lists, and
workshops to distribute the results and to educate researchers both
about the algorithms and the use of the applications.
\item {\bf Improved reproducibility} of numerical results by
publishing source codes used to obtain published results and by integration with the VisTrails~\cite{vistrails} provenance enabled workflow system.
\end{itemize}

In contrast to other fields, where open-source development efforts started decades ago, we can directly build our design on recent developments in computer science, especially:
\begin{itemize}
\item XML\cite{xml} (eXtensible Markup Language) and HDF5\cite{hdf5} (Hierarchical Data Format version 5)  as portable data formats supported by a large number of standard tools
\item generic and object oriented programming in C++ to achieve flexible but still optimal codes
\item OpenMP\cite{openmp} and MPI\cite{mpi} for parallelization on shared memory machines, clusters, and high performance supercomputers.
\end{itemize}

\subsection{New features in version 2.0}

The specific new features in version 2.0 are:
 \begin{itemize}
\item CMake build system~\cite{cmake}, simplifying configuration and enabling ALPS to be built on Windows.
\item Binary installer packages for MacOS X and Windows.
\item Using the binary file format HDF5~\cite{hdf5} in addition to XML, which is much faster and needs less space.
\item More flexible and powerful evaluation tools using Python.
\item A revised and substantially faster version of the directed loop quantum Monte Carlo (QMC) code {\tt dirloop\_sse}, and two new applications: QMC solvers for dynamical 
mean field theory ({\tt dmft}) and a time-evolving block decimation code ({\tt tebd}) for the dynamics of one-dimensional quantum systems.
\item Integration with the VisTrails workflow provenance system~\cite{vistrails}.
\item An expanded set of tutorials, and completely new tutorials using Python and VisTrails.
 \end{itemize}

\section{Building and Installing ALPS} \label{sct:build_and_install}
\subsection{Build system}
ALPS 2.0 uses CMake~\cite{cmake}, which is more flexible and portable than the previously used {\tt autotools}. All aspects of the build process are controlled by CMake variables, which can be easily changed in several ways, including a graphical user interface. This allows the user a large degree of customization of the ALPS installation. Furthermore, CMake supports the automatic creation of binary installation packages and building ALPS on Windows -- two often requested features. A snapshot of the build instructions at the time of the release are included in the source distribution. 
\subsection{Building ALPS from source}

Details about building ALPS from source and the required tools and libraries are discussed in \ref{sec:cmake}. This is the recommended way of building ALPS on Unix and Linux systems or if one wants to use specific BLAS libraries or Python interpreters on MacOS or Windows. Updated instructions as operating systems change will be made available on the ALPS web page~\cite{alps}.

\subsection{Binary Installation Packages}

To simplify installation on Mac OS X and Windows, we provide CMake-generated binary installation packages for these operating systems. To take advantage of all features of ALPS we recommend the user to download and install a binary installer for VisTrails~\cite{vistrails} in addition to the ALPS installers. On both operating systems the PATH variable should be adjusted to include the directory containing the ALPS binaries. The Windows installer gives one the option to do so automatically. The default installation directory on Mac OS X is {\tt /opt/alps/bin} and on Windows {\tt C:$\backslash$Program Files$\backslash$ALPS$\backslash$bin} on 32-bit versions and {\tt C:$\backslash$Program Files (x86)$\backslash$ALPS$\backslash$bin} on 64-bit versions, respectively.

Source and binary installation packages for Mac OS X using macports~\cite{macports} and for Linux using Ubuntu and Debian package managers will be made available in the near future.

\subsection{LiveALPS}

For Linux we additionally provide an ISO disk image, called LiveALPS, of a full Linux distribution with ALPS preinstalled. 
Based on a remastered CD-release of Knoppix~\cite{knoppix} it is ready to use 
on a PC without special configuration. While it can be burned to a DVD we recommend using it with a USB flash drive for performance reasons.
This solution is especially useful for Linux users that want to try ALPS, in summer schools, or in lectures.

\section{Data formats}

The most fundamental part of the ALPS project is the definition of
common standardized file formats suitable for a wide range of
applications. Standardized file formats enable the exchange of data
between applications, allow the development of common evaluation
tools, reduce the learning curve, simplify the application of more than one algorithm to a given
model, and are a prerequisite for the storage of simulation data in a
common archive.

\subsection{XML}

 The ISO
standard XML~\cite{xml} was chosen in ALPS version 1~\cite{ALPS1.2,ALPS1.3} for the specification of these formats
because it has become
the main text-based data format on the Internet and because it is
supported by a large and growing number of tools.

We have designed a number of XML  schemas \cite{xmlschema} to describe
\begin{itemize}
\item the input of simulation parameters,
\item the lattices,
\item quantum lattice models,   
\item and the output of results.
\end{itemize}

An introduction to XML and the ALPS XML formats is given in \ref{sec:xml}, and a  detailed specification of the formats is provided on
our web pages \cite{alps,xmlschema}.

\subsection{HDF5}

We now complement XML by the widely used Hierarchical Data Format 5 (HDF5)~\cite{hdf5}, a de-facto standard for writing large binary files in a portable, machine-independent way. HDF5 is directly supported by many visualization and data analysis tools and a       wide array of tools are 
available for C, C++, Fortran, Python, and other languages. In ALPS, HDF5 is now used as the default output format for simulation data. This approach is significantly faster than writing to text-based formats when large files need to be written. On-the-fly compression of binary data further reduces file size. 

The ALPS HDF5 files store simulation parameters, the detailed results of Monte Carlo simulations, spectra and expectation values of exact diagonalization and DMRG simulations, and time evolutions in the TEBD code. ALPS comes with a library of C++ and Python functions to load these results, but they can also be read with any other tool supporting HDF5. The exact schema of the HDF5 files is available on the ALPS web page~\cite{alps}. For backward compatibility with users' evaluation tools, the simulation codes accept a {\tt --write-xml} command line option to also write all results in XML.

The implementation of HDF5 capabilities within ALPS is split into two layers. The lower layer implements a generic and general-purpose C++ interface to the C HDF5 library. This layer is not tied to other parts of the ALPS library and can therefore be used by user code that is otherwise independent from ALPS and its file formats. The functions of this library are also exposed to Python, and can thus be used not only from C++ codes but also from Python scripts. On top of that, the specific file formats used by the ALPS library and applications are implemented in the respective components of the ALPS library.

\section{ALPS Libraries}
\noindent {\it Main developers:} L. Gamper, E. Gull. A. Hehn, B. Surer, S. Todo and M. Troyer. 

\medskip

The ALPS libraries are the foundation of all the ALPS applications, providing the functionality common to all of them:
\begin{itemize}
\item an XML parser and output stream to read and write the ALPS XML files.
\item a HDF5 library to read and write HDF5 files
\item an expression library to manipulate and evaluate symbolic expressions.
\item a scheduler for the automatic parallelization of Monte Carlo simulations and other embarrassingly parallel applications, implementing load balancing and checkpointing \cite{palm}.
\item a lattice structure library for the creation of arbitrary graphs and Bravais lattices from XML input.
\item a model library for the construction of basis sets, operators and Hamiltonians from XML input.
\item the ``alea'' library for the statistical analysis and evaluation of Monte Carlo data including a binning analysis of errors and jackknife analysis of cross-correlations \cite{palm}.
\item a serialization library ``osiris'' for the serialization of C++ data structures, used for writing program checkpoints and portable binary result files \cite{palm}.
\end{itemize}
These libraries make full use of object oriented and generic\cite{CE} programming techniques, thus being flexible while still not losing any performance compared to FORTRAN programs. To give just one example, the lattice library, implemented generically using C++ template features, is based on the Boost Graph Library\cite{BGL,boost} (BGL) and does not restrict the user to a specific data structure as in C or Fortran libraries. Instead the application programmer can choose the data structure best suited for the application, and as long as the data structure provides the BGL graph interface, the ALPS lattice library can construct the lattice from the XML description.

\section{Evaluation Tools in Python} \label{sct:python}
\noindent {\it Main developers:} B. Bauer, L. Gamper, J. Gukelberger, P.N. Ma, O. Parcollet, B.~Surer, M.~Troyer, and M.~L.~Wall. 

\medskip

The previous versions of ALPS had very limited data evaluation capabilities that were restricted to extracting plots from collections of XML files using XSLT. This is remedied in ALPS 2.0 by basing data evaluation on the powerful Python language~\cite{python}. Python is an easy-to-learn, interpreted, object-oriented language allowing interactive analysis of data and arbitrary complex evaluations.  Some brief examples of the Python language as well as the pyalps Python package new to ALPS 2.0 can be found in Appendix B.

We provide a complete set of library functions to write and read ALPS files and a number of useful functions to evaluate the simulation results and to make plots. In particular, the ALPS classes for the recording and evaluation of Monte Carlo data are all exported from C++ to Python, enabling an easy binning analysis~\cite{Ambegaokar2010} and jackknife analysis of Monte Carlo data.
Two-dimensional plots are created using the widely available Python matplotlib package~\cite{matplotlib} or by using ALPS functions to write grace~\cite{grace} or gnuplot~\cite{gnuplot} input files. Creation of 3D graphs is supported through VisTrails and the Visualization ToolKit (VTK)~\cite{vtk} VisTrails modules.

\subsection{The Python language}

Python~\cite{python} is an open-source, high-level, object-oriented interpreted programming language developed with a focus on portability and high productivity. This is achieved with a clear syntax and extensive standard libraries. A wealth of material is available for learning the language, including its own extensive standard documentation and several books\cite{LearningPython, ProgrammingPython}.

Python is an interpreted language, meaning that the source code is executed without an intermediate compilation step. This allows for interactive use, and easy analysis of simulation data. It also is simplifies code development and exchange, in particular since using an interpreted language also often improves the portability of the resulting code.

In addition to the Python standard library, the PyLab environment, which consists in particular of the NumPy~\cite{numpy}, SciPy~\cite{scipy} and matplotlib~\cite{matplotlib} libraries, offers powerful tools for numerical calculations and plotting. This makes Python particularly suitable for the data analysis part of the overall ALPS workflow. Wherever possible, the data types used in the pyalps package are built on top of the NumPy data types, allowing direct use of the PyLab libraries.

Python programs can interoperate with code written in a different language. As an example, a low-level numerical method could be implemented in C/C++ for performance reasons but still be called from a high-level Python code. This can be significantly simplified by using the Boost.Python~\cite{boost} library. Specifically, this library provides a simple interface to export C++ functions and classes to Python while taking care of Python internals such as memory management. We make use of this library to export parts of the ALPS library to Python in the pyalps package.

\subsection{The pyalps package}
ALPS 2.0 contains a package of Python modules for interfacing the low-level ALPS code to Python and vice versa, running simulations, and evaluation and post-processing of data. This package is known as pyalps. The advantages of using Python versus exclusively using low-level code for the above tasks are manifold. For example, Python's large standard library and ability to easily interface with low-level code allow for great flexibility in post-processing of data. Also, the use of Python facilitates use of the VisTrails workflow provenance tool, as VisTrails is written in Python.

Pyalps includes functions to write the xml files read by most applications, and additional functions to write special input files for applications with non-xml interfaces such as {\tt dmft} and {\tt tebd}. It also includes functions to run applications with specified input files both in series and in parallel. An example of the preparation of input files and running a simulation using pyalps is shown in Fig.~\ref{fig:python}.

Once the simulation has been run we must extract and evaluate data from it. At the heart of the pyalps evaluation tools is the {\tt DataSet} class. A {\tt DataSet} has as members {\tt x} and {\tt y} which are NumPy arrays containing numerical data. A {\tt DataSet} also has a member {\tt props} which is a dictionary -- one of the built-in types of Python -- of metadata describing the numerical data. The presence of the metadata allows for complex evaluations, for example when dealing with many {\tt DataSet}s from a parameter scan.

Using Boost.Python, large parts of the ALPS alea library for Monte Carlo observables are exported to Python. This allows the user to write Monte Carlo applications in Python while profiting from the full functionality of the ALPS C++ libraries. In addition, this allows the definition of numerical data types in Python which carry full statistical information. These types allow calculations with the syntax of normal floating-point numbers, but will perform a full error analysis. We also provide functions to read HDF5 files into a {\tt DataSet}. Wrapper functions around matplotlib for plotting of data stored in {\tt DataSet}s help the user by, e.g., automatically setting axis labels or legends from the metadata.

\section{Applications}
\label{sec:applications}
In addition to common libraries, the ALPS project includes a number of ready-to-use applications implementing the most important unbiased
algorithms for quantum many-body systems. All applications from ALPS 1.3, with the exception of the single particle DMRG demonstration program, have been retained in ALPS 2.0. They continue to work in the same way, with minor bug fixes and patches for incompatibilities, and we have added a number of new applications in version 2.0.

The applications all
share the same file formats, simplifying their use, reducing the
learning curve, and enabling the easy investigation of a model with
more than one method. Tutorials on the use of the applications are
included with the sources that can be found on the ALPS web
page~\cite{alps}.

\subsection{Exact diagonalization}

The exact diagonalization codes have been optimized for flexibility and not the highest performance or the goal to reach the biggest system sizes.

The codes make use of conserved quantum numbers, such as particle number or magnetization, and use translation symmetry to reduce the Hilbert space dimension. This speeds up the calculations and allows larger systems to be calculated. In addition, by calculating the energy eigenvalues separately for each momentum, the momentum-resolved excitation spectrum can be calculated.

The user may also specify custom measurements of averaged or local site- and bond operators, as well as arbitrary 2-point correlation functions. As an example, for the inhomogeneous bosonic Hubbard model of figure \ref{fig:hubmodel}, one can specify the measurement of the average double occupancy, local density, density correlations, and Green's function by defining the following input parameters, which use the definitions of site and bond operators provided with the model:

\begin{verbatim}
MEASURE_AVERAGE[Double] = double_occupancy
MEASURE_LOCAL[Local density] = n
MEASURE_CORRELATION[Density correlation] = n
MEASURE_CORRELATION[Green function] = "bdag:b"
MEASURE_STRUCTURE_FACTOR[Density Structure Factor] = n
\end{verbatim}

\subsubsection{Sparse diagonalization code {\tt sparsediag},} main developer: M. Troyer

\smallskip

\noindent  {\tt sparsediag} calculates the ground state and low lying excited states of quantum lattice models using the
Lanczos~\cite{lanczos} algorithm.

\subsubsection{Full diagonalization code {\tt fulldiag},} main developer: M. Troyer

\smallskip

\noindent {\tt fulldiag} calculates the complete
spectrum of quantum lattice models and from it all thermodynamic
properties. Access to the complete spectrum allows the calculation of all thermodynamic properties and of the temperature dependence of any measurement specified by the user.

\subsection{Classical Monte Carlo simulations} 

\subsubsection{Classical Monte Carlo codes for spin models {\tt spinmc},} main developer: M. Troyer

\smallskip

\noindent {\tt spinmc} is a classical Monte Carlo code for classical magnets employing local and cluster
updates~\cite{Swendsen87,Wolff89}. Supported models are Ising, XY, Heisenberg, and Potts models with isotropic and anisotropic interactions and magnetic fields.

\subsection{Quantum Monte Carlo simulations} 

\subsubsection{The looper code,} main developer: S. Todo

\smallskip

\noindent The {\tt loop} program implements the loop cluster
algorithm~\cite{Evertz93,Todo01,looper, Evertz03}, a generalization of the classical cluster updates \cite{Swendsen87,Wolff89} to quantum systems in both a path integral and a stochastic series expansion (SSE) representation \cite{Sandvik91,Sandvik99}. It supports isotropic and anisotropic models of quantum magnets in transverse and longitudinal fields.

\subsubsection{The directed loop algorithm code {\tt dirloop\_sse},} main developer: S.~Isakov and L.~Pollet

\smallskip

\noindent The {\tt dirloop\_sse} code in ALPS 2.0 is a completely new and improved implementation of the directed loop QMC algorithm~\cite{Sylyuasen,Alet2005}, replacing the previous one in ALPS 1.2 and 1.3.  It is ideal for spin models in magnetic fields, frustrated spin models, and hard-core boson models.
 
\subsubsection{The worm algorithm code {\tt worm},} main developers: S. Trebst and M. Troyer

\smallskip

\noindent The {\tt worm} code implements the worm algorithm~\cite{Prokofev98A} and is ideal for bosonic models.

\subsubsection{The extended ensemble code {\tt qwl},} main developers: M. Troyer and S. Wessel

\smallskip

\noindent The {\tt qwl} code is an extended ensemble QMC program using a generalization of the Wang-Landau\cite{Wang01,Wang01b} algorithm to quantum systems,\cite{Troyer03} to obtain thermodynamic quantities over large temperature ranges.

\subsection{Density Matrix Renormalization Group (DMRG) algorithms} 

\subsubsection{The {\tt dmrg} code,} main developers: A. Feiguin (core code) and M. Troyer (ALPS interface)

\smallskip 

The {\tt dmrg} code implements the DMRG algorithm~\cite{White1992,Schollwock2005} to calculate ground states and low-lying excited states of quasi-one dimensional quantum systems. It can also measure arbitrary local quantities and two-point functions.

\subsubsection{Time Evolving Block Decimation Code {\tt tebd},} main developers: L.~D.~Carr and M.~L.~Wall

\smallskip

Release 2.0 contains interfaces to the Open Source TEBD project~\cite{ostebd}, which is a collection of Fortran libraries using the Time-Evolving Block Decimation (TEBD) algorithm~\cite{vidal1, vidal2} to simulate time evolution of one-dimensional quantum systems. TEBD can also find ground states via imaginary time evolution. The TEBD routines included in ALPS are an updated version of the v2.0 release of Open Source TEBD with improvements for speed and numerical stability.

At present, the TEBD routines can simulate the spin, boson Hubbard, hardcore boson, spinless fermion, and fermion Hubbard models from the ALPS models library. All 
Hamiltonian parameters are assumed uniform throughout the system. Because TEBD produces wavefunctions, a wide array of observables can be computed including local quantities, two-point correlation functions, entanglement measures, and overlaps between the state at different times. Observables calculated are the $z$ and $x$ magnetizations, their squares, and the $\langle \hat{S}^z_i 
\hat{S}^z_j\rangle$ and $\langle \hat{S}^x_i \hat{S}^x_j\rangle$ correlations for the spin model, the occupation number, its square, and the $\langle \hat{n}_i \hat{n}_j\rangle$ and 
$\langle \hat{a}_i^{\dagger} \hat{a}_j\rangle$ correlation functions for the boson Hubbard, hardcore boson, spinless fermions, and fermion Hubbard models; and, additionally, 
the magnetization and $\langle \hat{S}^z_i \hat{S}^z_j\rangle$ correlation function for the fermion Hubbard model. All models calculate the energy, von Neumann entanglement 
entropy of each site, the von Neumann entanglement entropy on each bond, and the overlap of the wavefunction at time $t$ with the state at $t=0$ ({\it i.e.}, the Loschmidt echo).

\subsection{Dynamical Mean Field Theory QMC Solvers {\tt dmft}}

\noindent {\it Main developers: } S. Fuchs, E. Gull, B. Surer, M. Troyer and P. Werner

\medskip

Dynamical mean field theory (DMFT) is a method to simulate fermionic lattice systems which approximates the self-energy $\Sigma(k,\omega)$ by a momentum-independent 
function $\Sigma(\omega)$. For such a ``local'' self-energy, the diagrammatic structure simplifies considerably and the lattice problem can be mapped onto an impurity problem subject to a self-consistency condition for the bath~\cite{Georges96,Kotliar06}. The method may be extended from a single-site to a cluster formalism, thereby rendering it controlled in practice~\cite{Maier05}.
We provide a simple implementation of the DMFT self-consistency loop for general single-site multi-orbital problems.
For the solution of the quantum impurity model we provide a legacy Hirsch-Fye QMC code~\cite{Hirsch86} as well as two continuous-time QMC algorithms.

The ``interaction expansion'' algorithm~\cite{Rubtsov04,Rubtsov05} expands the impurity model partition function in powers of the interaction terms and samples the resulting diagrams stochastically. This method, historically
the first continuous-time quantum Monte Carlo impurity solver algorithm, is particularly suitable for the simulation of impurity clusters.
The complementary hybridization expansion algorithm~\cite{Werner06,Werner06Kondo} expands the impurity model partition function in the impurity-bath hybridization, treating 
the local (impurity) Hamiltonian exactly. 
The method is useful to simulate multi-orbital models, since it can easily treat general interaction terms. 
The current implementation of the ALPS code (both for the interaction expansion and the hybridization expansion) is restricted to (multiorbital) single site and density-density interactions.
A detailed description of the ALPS DMFT algorithms is given in Ref.~\cite{ALPSDMFT}.

\section{Integration with the VisTrails Workflow and Computational Provenance Tools}
\noindent{\it Main developers:} B. Bauer, J. Freire, L. Gamper, J. Gukelberger, D. Koop, P. Mates, E. Santos, V.W. Scarola, C. Silva, B. Surer and M. Troyer.

\subsection{Computational Provenance}

ALPS 2.0 seeks to ensure result reproducibility. Provenance (also
referred to as history, audit trail, lineage, or pedigree) captures
information about the steps used to generate a given result
\cite{Silva07,Freire08}. Such information is crucial in data
preservation and determining data quality as well as interpreting,
reproducing, sharing, and publishing results. Release 2.0 improves
upon previous ALPS versions by using the VisTrails~\cite{vistrails}
workflow system to record provenance-related information, including
algorithm workflows, simulation parameters, and version history to automate
reproducibility.

VisTrails is an open-source system that was designed to support
exploratory computational tasks such as visualization and data mining
while ensuring computational provenance~\cite{vistrails,Bavoil05}.
VisTrails enforces provenance with workflows that allow a capture
mechanism and an infrastructure for storage, access, and queries. In Appendix A we present a concise introduction to scientific workflows and VisTrails.

\begin{figure}
\begin{center}
\href{http://arxiv.org/src/1101.2646/anc/7e766f827f1bad8c8df574a0cc6135ce.vtl}
{\includegraphics[width=8cm]{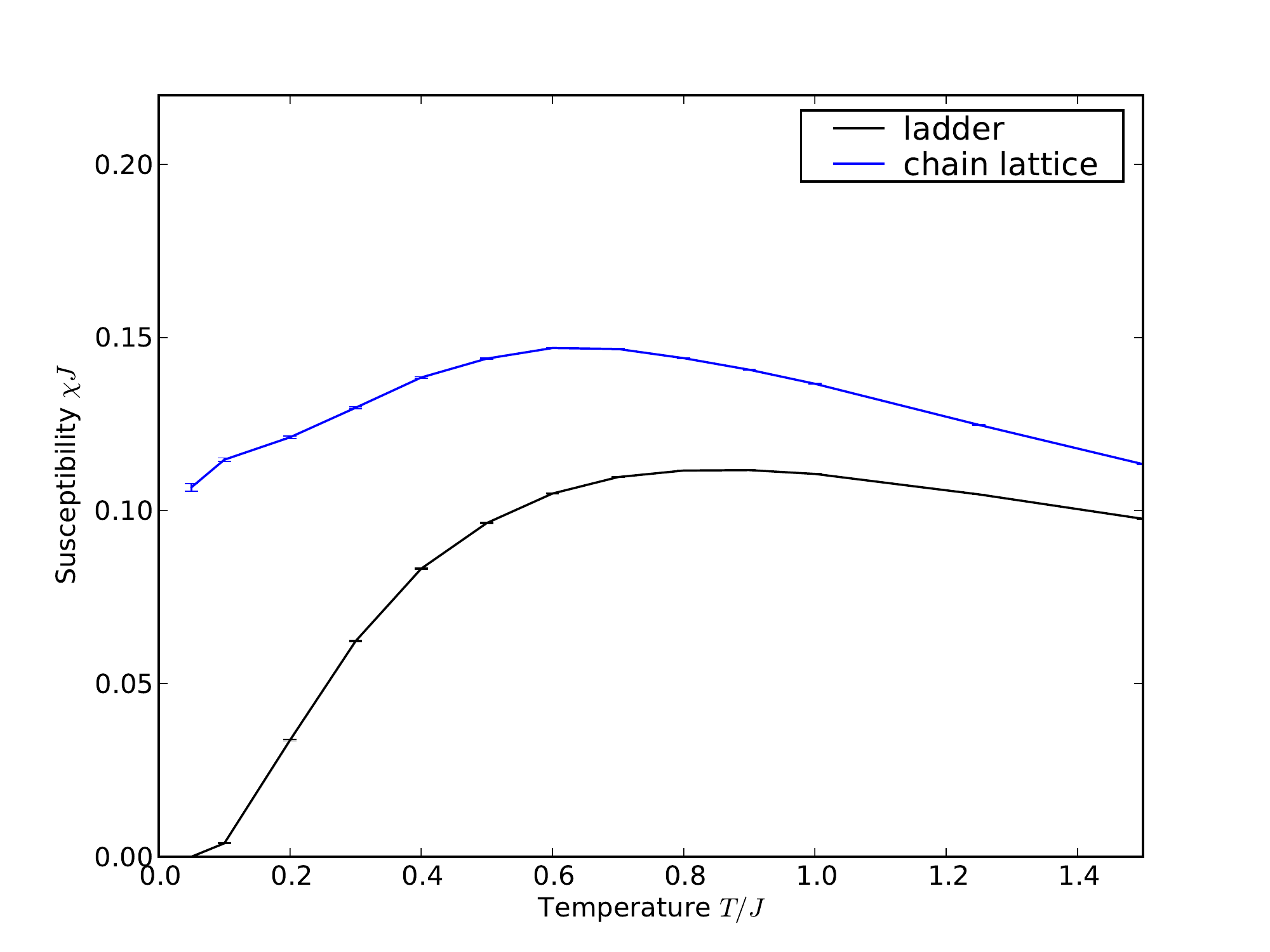}}
\caption{A figure produced by an ALPS VisTrails workflow: the uniform susceptibility of the Heisenberg chain and ladder. Clicking the figure retrieves the workflow used to 
create it. Opening that workflow on a machine with VisTrails and ALPS installed lets the reader execute the full calculation.}
\label{fig:figure}
\end{center}
\end{figure}

\begin{figure}
\begin{center}
\href{http://arxiv.org/src/1101.2646/anc/7e766f827f1bad8c8df574a0cc6135ce.vtl}
{\includegraphics[width=8cm]{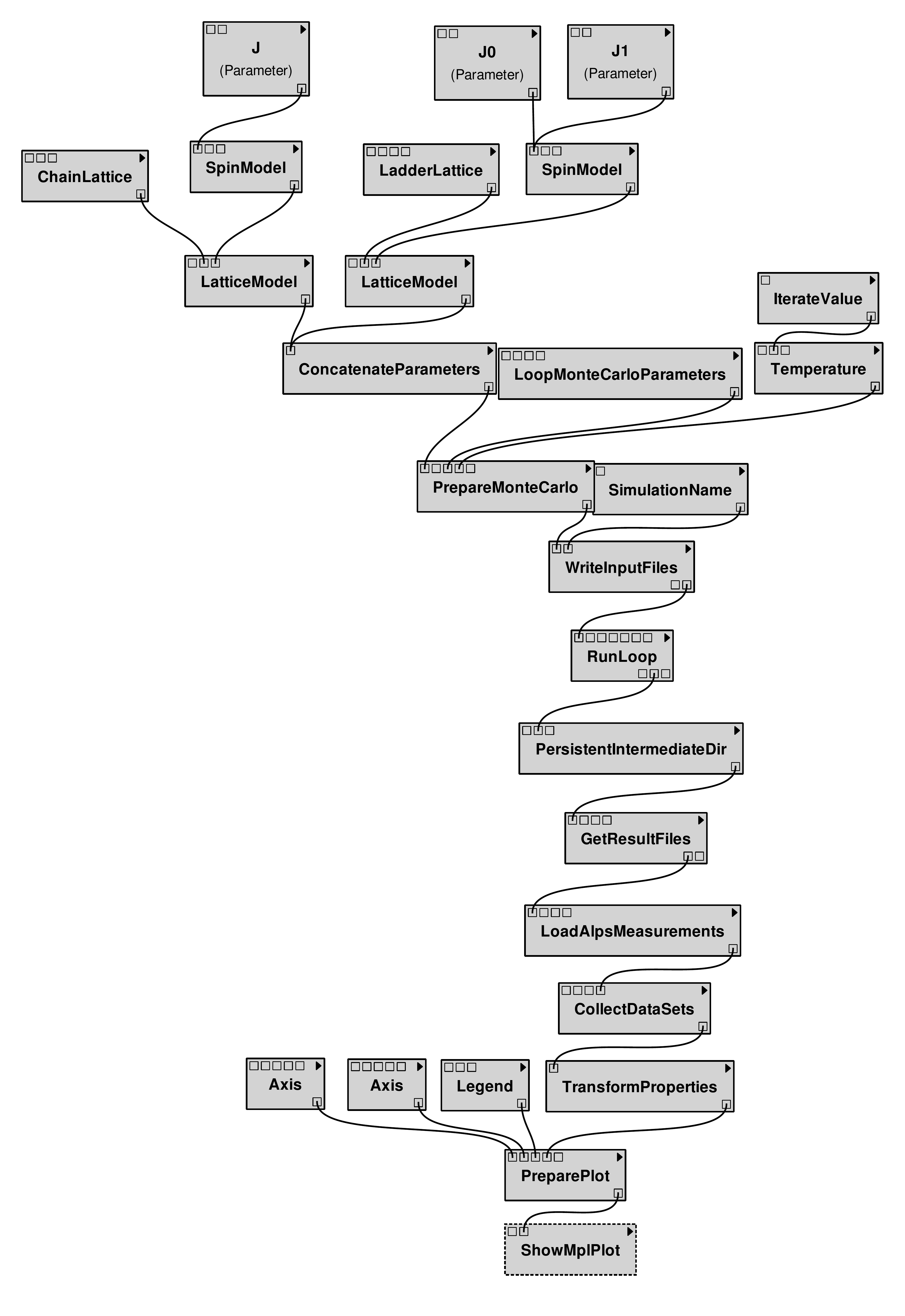}}
\caption{The workflow that created Fig.~\ref{fig:figure}. The workflow image has been created by VisTrails for the specific workflow used to create the exact version shown in the figure. Clicking the figure retrieves the workflow.}
\label{fig:workflow}
\end{center}
\end{figure}

\subsection{The ALPS VisTrails package}

ALPS 2.0 provides a package of modules for VisTrails, built on top of the pyalps package, that allows the user to perform the three major steps of a scientific workflow, namely i) the preparation of input files, ii) the execution of simulations, iii) the analysis and plotting of simulation data. The three groups of modules are intended to integrate seamlessly, but at the same time allow the separate use of each group, which for example allows the user to analyze data with VisTrails that has not been prepared using the ALPS library or file formats.

The design goal of the data analysis modules is to take care of common, repetitive tasks, while at the same time being as flexible as scripts. Common data formats, designed to allow interoperability with the powerful NumPy/SciPy libraries, are used by all modules. A set of self-contained modules are provided for common tasks, such as fitting or various types of plots. A framework of {\it Transform} modules allows the user to specify arbitrary transformations by providing small Python codes.

The modules for input preparation and data analysis are designed to assist the user in dealing with a large number of simulations. For example, modules are provided to sweep parameter ranges, to combine data from many simulations or to load data from ALPS files and text files. Modules to load data from custom file formats can also easily be implemented.

VisTrails workflows in ALPS 2.0 use descriptive nouns and verbs for the names of modules that execute
 Python scripts which in turn execute ALPS
algorithms in Python or compiled code. Fig.~\ref{fig:workflow} shows an example ALPS workflow
that takes input lattice parameters and model definitions (modules described by nouns, such as 
 ChainLattice and LatticeModel), executes the quantum
Monte Carlo looper code, and then renders the data (modules described by verbs such as RunLoop and PreparePlot).

The creation of such workflows is explained in a set of tutorials and on the web page. Furthermore, all ALPS tutorials, discussed in the next section, are available as VisTrails workflows, providing an extensive set of examples that can be used as templates for the user's simulations. 

This paper already makes use of ALPS 2.0 and VisTrails. Full provenance information for the results in Fig.~\ref{fig:figure} is available from the URL linked to the figure. Clicking the figure will download the workflow that has been used to prepare the figure and its version history. After installing VisTrails and ALPS, the reader of the paper can redo the full simulation.

\begin{figure}[t]
\centering
\href{http://arxiv.org/src/1101.2646/anc/291afe8abedfc9c7303b6b5fa832309a.vtl}
{\includegraphics[width=8cm]{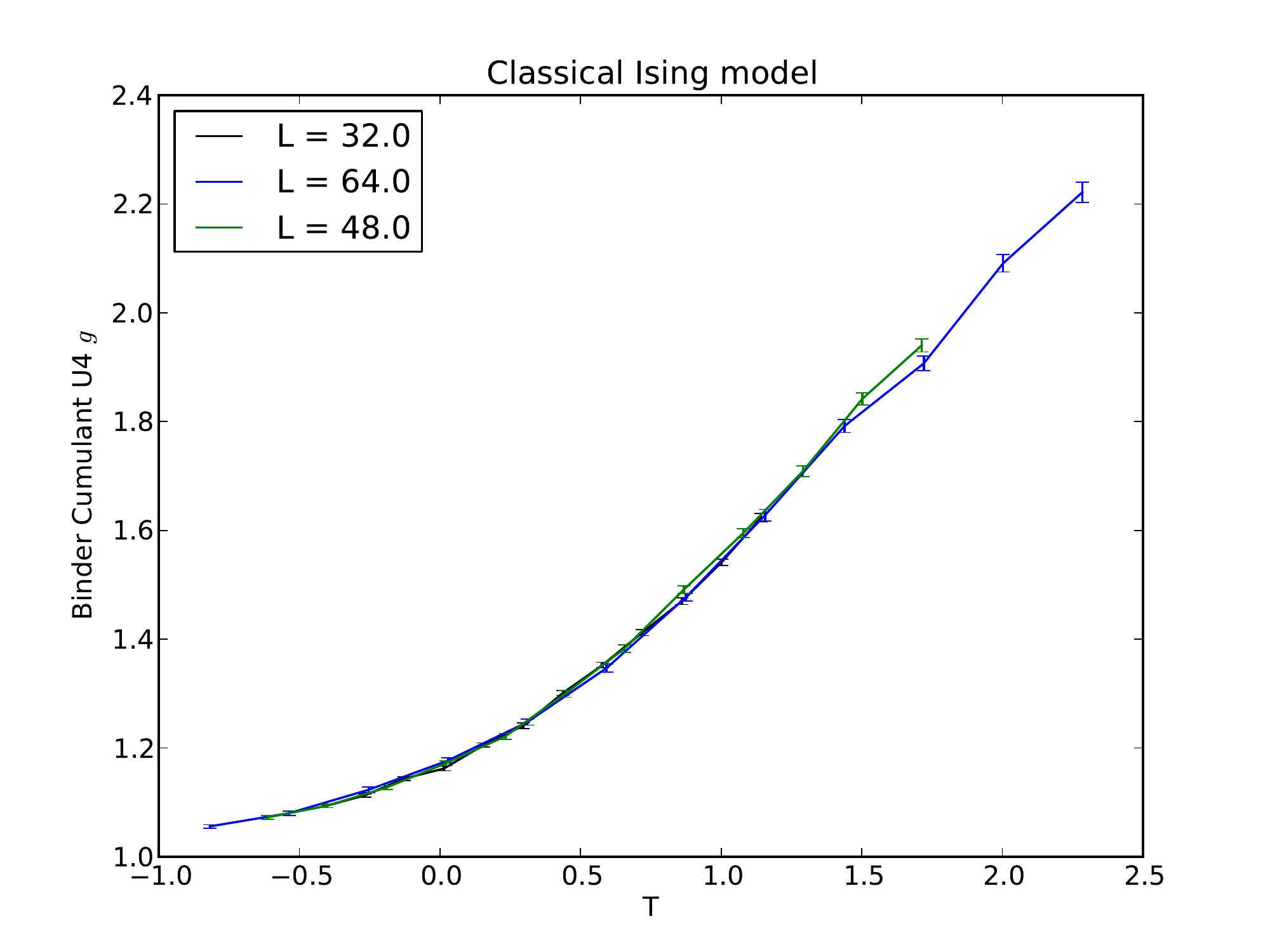}}
\caption{In this example we show a data collapse of the Binder Cumulant in the classical Ising model. The data has been produced by remotely run simulations and the critical exponent has been obtained with the help of the VisTrails parameter exploration functionality.}
\label{fig:datacollapse}
\end{figure}

ALPS VisTrails is also suitable for data analysis only. For compute-intensive problems requiring computer cluster resources the data analysis within Vistrails is performed by reading the observables from stored result files. Operating in this way is the workflow producing Fig.~\ref{fig:datacollapse} showing a data collapse of the Binder cumulant in the classical Ising model. The workflow downloads the corresponding result files from a server, loads the Binder Cumulant observable and allows to interactively determine the critical exponents by a parameter exploration. The reader is invited to explore the sensitivity of the data collapse to the correlation length critical exponent.

\begin{figure}[t]
\begin{multicols}{2}
\begin{lstlisting}
cat > parm << EOF
LATTICE="chain lattice"
MODEL="spin"
local_S=1/2
L=60
J=1
THERMALIZATION=5000
SWEEPS=50000
ALGORITHM="loop"
{T=0.05;}
{T=0.1;}
{T=0.2;}
{T=0.3;}
{T=0.4;}
{T=0.5;}
{T=0.6;}
{T=0.7;}
{T=0.75;}
{T=0.8;}
{T=0.9;}
{T=1.0;}
{T=1.25;}
{T=1.5;}
{T=1.75;}
{T=2.0;}
EOF

parameter2xml parm
loop --auto-evaluate --write-xml parm.in.xml
\end{lstlisting}
\end{multicols}
\caption{A shell script to perform an ALPS simulation to calculate the uniform susceptibility of a Heisenberg spin chain. Evaluation options are limited to viewing the output files. Any further evaluation requires the use of Python, VisTrails, or a program written by the user.}
\label{fig:commandline}
\end{figure}
\section{Tutorials and Examples}

\noindent{\it Main contributors: } B. Bauer, A. Feiguin, J. Gukelberger, E. Gull, U. Schollw\"ock, B.~Surer, S. Todo, S. Trebst, M. Troyer, M.L. Wall and S. Wessel

\medskip

The ALPS web page~\cite{alps}, which is a community-maintained wiki system and the central resource for code developments, 
also offers extensive resources to ALPS users. In particular, the web pages feature an extensive set of tutorials, which for each ALPS application 
explain the use of the application codes and evaluation tools in the context of a pedagogically chosen physics problem in great detail.
These application tutorials are further complemented by a growing set of tutorials on individual code development based on the ALPS libraries.

\subsection{Tutorials on Using the ALPS Codes}
We provide tutorials and instructions for using ALPS in three ways: 
\begin{enumerate}
\item from the command line (without data evaluation and plotting)
\item using Python scripts
\item and using the VisTrails workflow system. 
\end{enumerate}

The input files needed for the tutorials are available on the web page but are also included in the source and binary distributions.

\begin{figure}
\begin{lstlisting}
import pyalps
import matplotlib.pyplot as plt
import pyalps.plot

#prepare the input parameters
parms = []
for t in [0.05, 0.1, 0.2, 0.3, 0.4, 0.5, 0.6, 0.7, 0.8, 0.9, 1.0, 1.25, 1.5, 1.75, 2.0]:
 parms.append(
  { 
   'LATTICE'  : "chain lattice", 
   'MODEL'   : "spin",
   'local_S'  : 0.5,
   'T'    : t,
   'J'    : 1 ,
   'THERMALIZATION' : 5000,
   'SWEEPS'   : 50000,
   'L'    : 60,
   'ALGORITHM'  : "loop"
  }
 )

#write the input file and run the simulation
input_file = pyalps.writeInputFiles('parm2c',parms)
pyalps.runApplication('loop',input_file)

#load the susceptibility and collect it as function of temperature T
data = pyalps.loadMeasurements(pyalps.getResultFiles(prefix='parm2c'),'Susceptibility')
susceptibility = pyalps.collectXY(data,x='T',y='Susceptibility')

#make plot
plt.figure()
pyalps.plot.plot(susceptibility)
plt.xlabel('Temperature $T/J$')
plt.ylabel('Susceptibility $\chi J$')
plt.title('Quantum Heisenberg chain')
plt.show()
\end{lstlisting}
\caption{A Python script to perform an ALPS simulation to calculate the uniform susceptibility of a Heisenberg spin chain, load and evaluate the data, and make a plot. }
\label{fig:python}
\end{figure}

An example for running an ALPS application from the command line is shown in Fig.~\ref{fig:commandline}. This example calculates the uniform susceptibility of a quantum Heisenberg spin chain. The same example, containing additional code to evaluate and plot the results using Python is shown in Fig.~\ref{fig:python}. The workflow using VisTrails is shown in Fig.~\ref{fig:workflow}, and its version history in Fig.~\ref{fig:history}. Following the links in the PDF version of the latter figures retrieves the corresponding vistrail files.

\subsection{Tutorials on Writing Codes with ALPS}

A second set of tutorials demonstrate how to use ALPS libraries and tools in new simulation codes. We provide examples of how to write a simple Monte Carlo simulation in either Python or C++ and how to integrate it with the ALPS tools. We also demonstrate how a user's code can be easily integrated into the VisTrails workflow system.

\section{License}
The ALPS libraries are licensed under the ALPS library license~\cite{librarylicense} and the applications under the ALPS application license~\cite{applicationlicense}. These licenses are modeled after the GNU General Public License (GPL), but contain an additional requirement to cite this paper as well as relevant papers for each of the application codes. These papers need to be cited and the use of ALPS acknowledged in any scientific project that makes use of ALPS. This includes the case when ALPS has only been used to test a scientist's application code.

The detailed license text is included in the files {\tt LICENSE.txt}~\cite{librarylicense} and {\tt LICENSE-applications.txt}~\cite{applicationlicense}. Any use of ALPS requires citing this paper. These files can be found at the top level of the source distribution and in the directory {\tt share/alps} of the binary distributions. The papers that need to be cited for the use of a specific application are printed to the standard output when that application is run. The list of papers is also included in the file {\tt CITATIONS.txt}. Since some of those references are to preprints we recommend that the ALPS web page~\cite{alps} is checked for updates.

\section{Future Development plans}

The ALPS project is a work in progress and development will continue after this release 2.0. Immediate plans for release 2.1 within about a year include the development of more evaluation tools for Monte Carlo simulations and more 2D and 3D plotting functionality.

ALPS 2.1 will also provide support for more programming languages. In addition to the Python support in the current version we plan to add a Fortran library to write the ALPS HDF5 files from Fortran. This will enable the use of the ALPS evaluation tools not only with C++ or Python codes, but also with results of Fortran codes.

A key part of ALPS 2.1 will be a more flexible and optimized scheduler. It will allow the simple integration of ALPS codes into other programs and Python scripts. It is also designed to scale to tens of thousands of CPUs, compared to the current scheduler that scales only to a few thousand CPUs. 

Besides improved versions of some of the simulation codes, there will be a new worm algorithm program for optical lattice simulations with millions of lattice sites, and an extension of the DMFT codes to clusters.

ALPS is an open initiative and we welcome contributions from the community.

\section{Acknowledgements}

We thank A.F. Albuquerque, F. Alet, P. Corboz, P. Dayal, A. Honecker, A. L\"auchli, M. K\"orner, A. Kozhevnikov, S. Manmana, M. Matsumoto, I.P. McCulloch, F. Michel and R.M. Noack, for their contributions to previous versions of ALPS and for useful discussions. We thank D. Abrahams for developing Boost.Python and for his support of the Boost libraries.

We thank all users of the previous versions of ALPS and beta testers of ALPS 2.0, especially J. Alfonsi, S. Aicardi, A. Akande, P. Anders, A. Anfossi, Z. Asadzadeh, R.~Bhat, J.H. Brewer, S. Bulut, E. Burovski, G. Carleo, G. Chen, M.D. Costa, M.~Dolfi, M.~Ferrero, J. Figgins, S.~Gazit, U.~Gerber, R. Ghulghazaryan, C. Gils, S. Greschner, C.~Hamer, J. Hammond, K. Hassan, N. Heine, F.-J. Jiang, R. Jordens, H.G. Katzgraber, B.~Keiyh, J. Kim, J. Kolasinski, B.~Koopman, F. K\"ormann, E.~Kozik, M.F. Krenn, F.~Kruger, M. Laurent, J. Lopes, M. Maik, M. Maksymenko, N. Moran, A. Nunnenkamp, J.D. Peel, S.~Pilati, K. Prsa, N.~Rahman, M. Rofiq, H. R\o nnow, J. Ruostekoski, J.~Schnack, N. Schuch, A. Sen, K. Shtengel, J. Single, M. Skoulatos, C. Sliwa, M.~Spenke, M.~Stoudenmire, V. Tangoulis, A.~Taroni, B. Thielemann, M. Tovar, H. Tran, V.~Varma, S. Ward, T. Wasiutynski, J. Wen, J. Wilms, H. Wunderlich, F. Xiao, X.Q. Yu, J.~Zakrzewski, Z.-X. Zhou, and R. Ziemkiewicz. We thank P. Zoller for the suggestion to provide binary installers. 

The development of ALPS has profited from support of the Pauli Center at ETH Z\"urich, the Swiss HP$^2$C initiative, the Kavli Institute for Theoretical Physics in Santa Barbara through NSF grants PHY-0551164 and DMR-0705847, the Aspen Center for Physics, the Swiss National Science Foundation, the Jeffress Memorial Trust, Grant No.~J-992, the National Science Foundation under grant PHY-0903457 and DMR-0955707, the Golden Energy Computing Organization at the Colorado
School of Mines for the use of resources acquired with financial assistance from the
National Science Foundation and the National Renewable Energy Laboratories, the Deutsche Forschungsgemeinschaft through the collaborative research center SFB 602, Japan Society for the Promotion of Science through KAKENHI No.\,20540364, Next-Generation Supercomputer Project from MEXT Japan, and a grant from the Army Research Office with funding from the DARPA OLE program.

Work on VisTrails is primarily supported by grants and contracts from
the U.S. National Science Foundation, the U.S. Department of Energy,
and IBM.

\appendix

\section{Scientific Workflows and VisTrails}

\subsection{Workflows}

Workflow systems provide well-defined languages for specifying complex
tasks from simpler ones; they capture complex processes at various
levels of detail and systematically record the provenance information
necessary for automation, reproducibility, and result sharing.

We start by introducing basic concepts underlying
workflows and provenance along with the basic terminology.
Scientific workflows are often used to perform data intensive tasks
and are represented as dataflow networks~\cite{lee@ieee1995} where the
execution order is determined by the flow of data through the
workflow. Scientific workflow systems offer a number of advantages
over programs and scripts for constructing and
managing computational tasks. In addition to providing a simple
programming model, many systems provide intuitive visual programming
interfaces which make them more usable for users who do not have
substantial programming expertise. As we discuss below, the structure
present in workflow definitions also enables a series of operations
and queries that simplify the manipulation and re-use of workflows.
From now on, we use the terms scientific workflow, workflow, pipeline,
and dataflow interchangeably.

\begin{figure}[t]
\includegraphics[width=\linewidth]{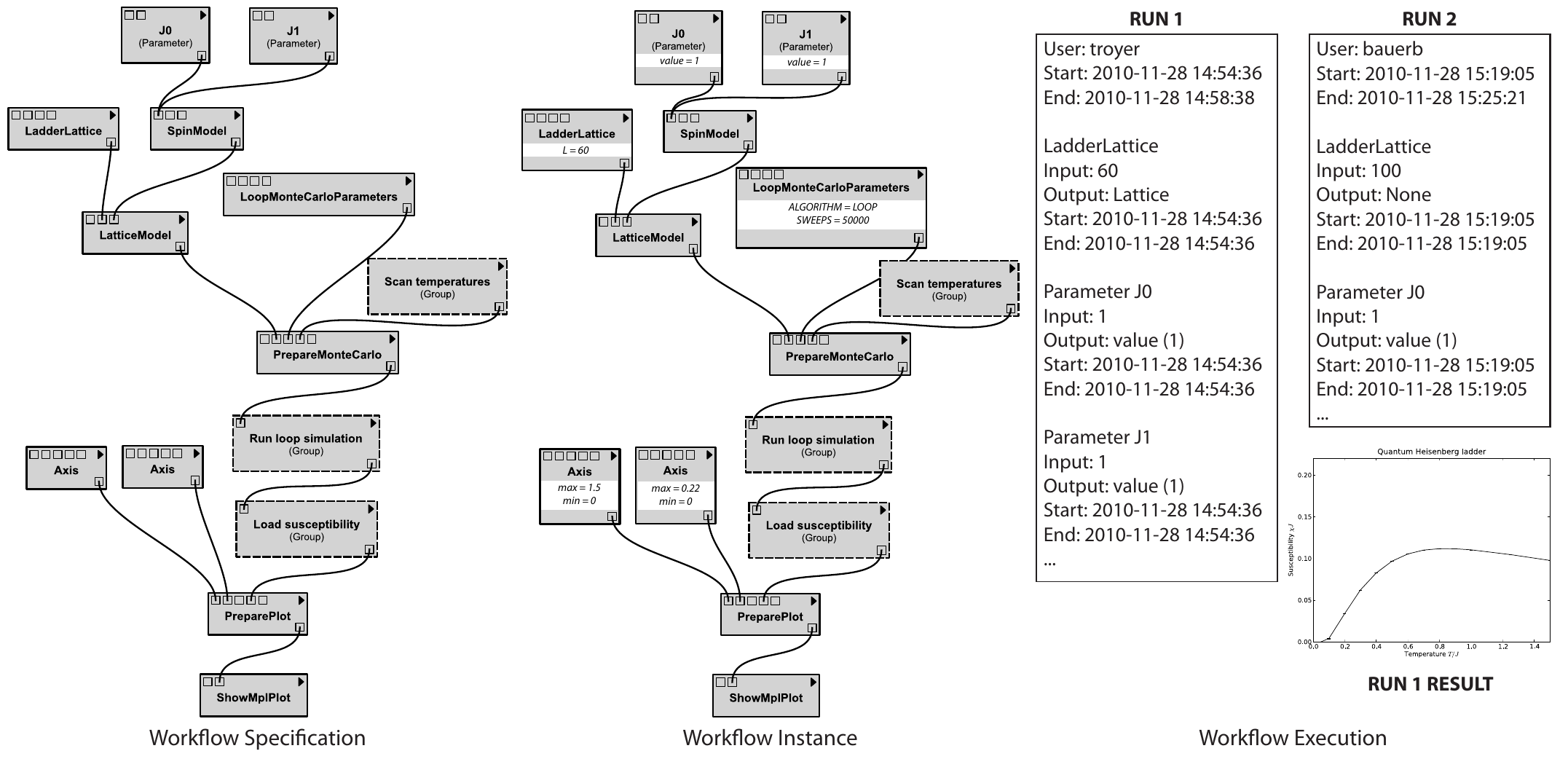}
\caption{The workflow specification (on the left) shows a workflow to calculate the uniform magnetic susceptibility of a quantum Heisenberg spin ladder and plot it as a function of temperature.
 The workflow
 instance (in the middle), in addition to the workflow specification,
 contains value assignments for parameters of the modules. This
 instance was used to derive the plot in the bottom right. The
 workflow runs (on the right) consist of information collected while
 the workflow instance was executed. }
\vspace{0cm}
\label{fig:workflow-spec-run}
\end{figure}

Formally, a \emph{workflow specification} is defined by a tuple
$(M,I,O,C,p:O \cup I \rightarrow M)$, where $M, I, O, C$ are,
respectively, sets of {\it modules}, {\it input ports}, {\it output
 ports}, {\it connections}, and $p$ is a function that assigns a
unique module to input and output ports. If $x \in I \cup O$ and $m =
p(x)$ then we say that $x$ {\it is a port of module} $m$ and that $m$
is {\it the module of} $x$. The set $C$ of connections is a subset of
$O \times I$. In a {\it connection} $(o,i) \in C$, $o$ is called the
{\it source port} and $i$ is called the {\it target
 port}. Furthermore, the module of the source port cannot be the same
as the module of the target port. Additionally each port (input
or output) has an associated type and a {\it name} that is unique
across the ports of the same module. A {\it module signature} is
defined as a set of pairs, where each pair contains a port name and
its type. In addition, each module has a set of {\it parameters}. Each parameter
has a unique {\it name} across the set of parameters of the same
module.

A \emph{workflow instance} consists of a specification combined with
bindings that provide values to parameters in the modules. A {\it
 workflow run} is the execution of the following algorithm for a
workflow instance: each module checks whether all its input ports that
are target ports of a connection have a value; if false it waits; if
true, the module, according to its semantics, produces output values
on its output ports; values that are produced in an output port flow
to target ports of all the connections that have that output port as
its source port; this algorithm continues until there are no more
modules to execute.
These concepts are illustrated in Figure~\ref{fig:workflow-spec-run}.

\subsection{Data Provenance}
In the context of scientific workflows, data provenance is a record of
the derivation of a set of results. There are two distinct forms of
provenance~\cite{VDL:Challenge06}: prospective and retrospective.
\emph{Prospective provenance} captures the specification of a
computational task (\ie a workflow specification or instance)---it
corresponds to the \emph{steps that need to be followed} (or a recipe)
to generate a data product or class of data products.
\emph{Retrospective provenance} captures the \emph{steps that were
 executed} (\ie the workflow run) as well as information about the
execution environment used to derive a specific data product---a
detailed log of the execution of a computational task.

An important piece of information present in workflow provenance is
information about \emph{causality}: the dependency relationships
between data products and the processes that generate them. Causality
can be inferred from both prospective and retrospective provenance.
Data provenance also includes \emph{user-defined information}, such as
documentation that cannot be automatically captured but records
important decisions and notes. This data is often captured in the
form of annotations.

\paragraph{Provenance of Workflow Evolution.}
Although workflows have been traditionally used to automate repetitive
processes, for many exploratory scientific tasks such as data analysis
and visualization, change is the norm. As users formulate and test
hypotheses, they often need to refine and compare the results of
several workflows. VisTrails~\cite{vistrails} supports a novel
change-based provenance model that treats workflow instances as
first-class data
products~\cite{Freire:2006:IPAW,callahan@sciflow2006}. As a scientist
makes modifications to a workflow, the system transparently records
those changes (\eg the addition of a module, the modification of a
parameter, \etc) akin to a database transaction log. Note that the
change-based model uniformly captures both changes to parameter values
and to workflow definitions. This sequence of changes is sufficient to
determine the provenance of data products, and it also captures
information about how the workflows evolve over time. We refer to
this detailed provenance of the workflow evolution as a visual trail,
or a \emph{vistrail}.

\begin{figure}[t]
\centering
\begin{center}
 \includegraphics[width = 1.0\linewidth,clip=false]{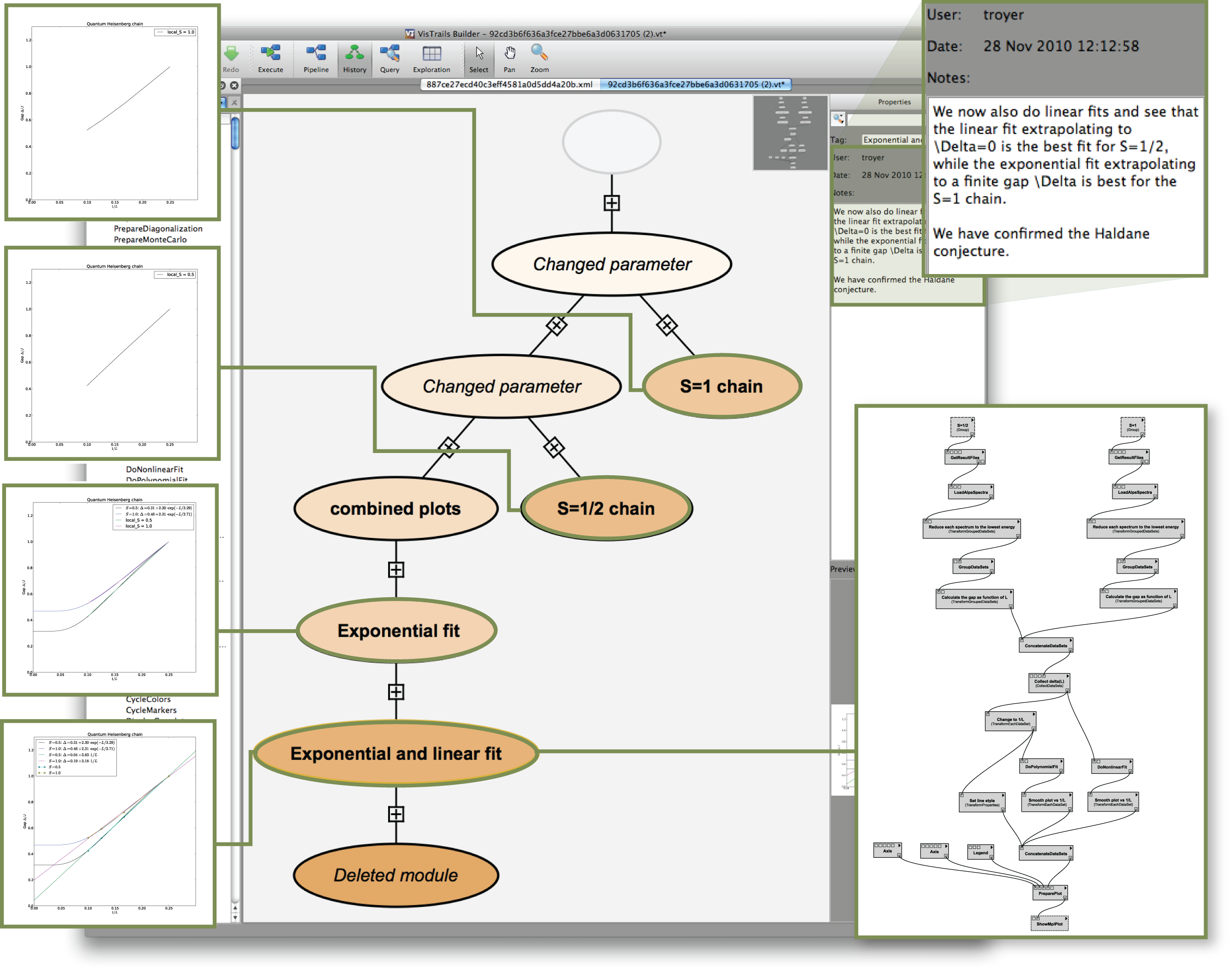}
 \caption{Workflow evolution provenance. 
 In this example we have calculated the system size dependence of the spin gap of spin-1 and spin-1/2 Heisenberg chains, first calculating the gap by exact diagonalization and then performing fits to $\Delta(L)= c/L$ and $\Delta(L)= \Delta_0 + c \exp(-L/\xi)$. As expected from the Haldane conjecture the plots show that the first fit is best for the gapless spin-1/2 chain and the second fit is best for the gapped spin-1 chain.
 Complete provenance of the exploration process is
 displayed as a history tree with each node representing a workflow
 instance that generates a plot. Detailed meta-data is
 also stored including free-text notes made by the scientist, the
 date and time the workflow was created or modified, an optional
 descriptive tag, and the user that created it.} \label{fig:alps-tree}
 \vspace{-.3cm}
\end{center}
\end{figure}

As Figure~\ref{fig:alps-tree} illustrates, each node in a vistrail
corresponds to a workflow instance and contains value assignments for
the module parameters. A series of workflow runs may be associated
with a node as well as additional metadata, such as user annotations.
A tree-based view of a vistrail allows a scientist to return to a
previous version in an intuitive way, to undo bad changes, to compare
different workflows, and to be reminded of the actions that led to a
particular result. Another important benefit of the change-based
provenance model is that it enables a series of operations that
simplify the exploration process, in particular, the ability to refine
workflows by analogy~\cite{analogies@tvcg2007}, as well as to visually
compare workflows and their results~\cite{Freire:2006:IPAW}.

It is important to note that the provenance of workflow execution and
workflow evolution help inform each other, leading to improved methods
for using the combined information. Knowing which workflow instances
have executed correctly can explain why workflows were updated, and
the number of times that an instance has been executed can help
highlight important versions of workflows. Similarly, knowing how a
workflow evolved can help users debug errors that occur when running
modified versions. Change-based provenance from workflow evolution
can also speed up runs of similar workflows via
caching~\cite{bavoil@vis2005}.

\begin{figure}
\begin{center}
\href{http://arxiv.org/src/1101.2646/anc/577d23d03ef2d80432809613e7f083ac.vtl}
{\includegraphics[width=8cm]{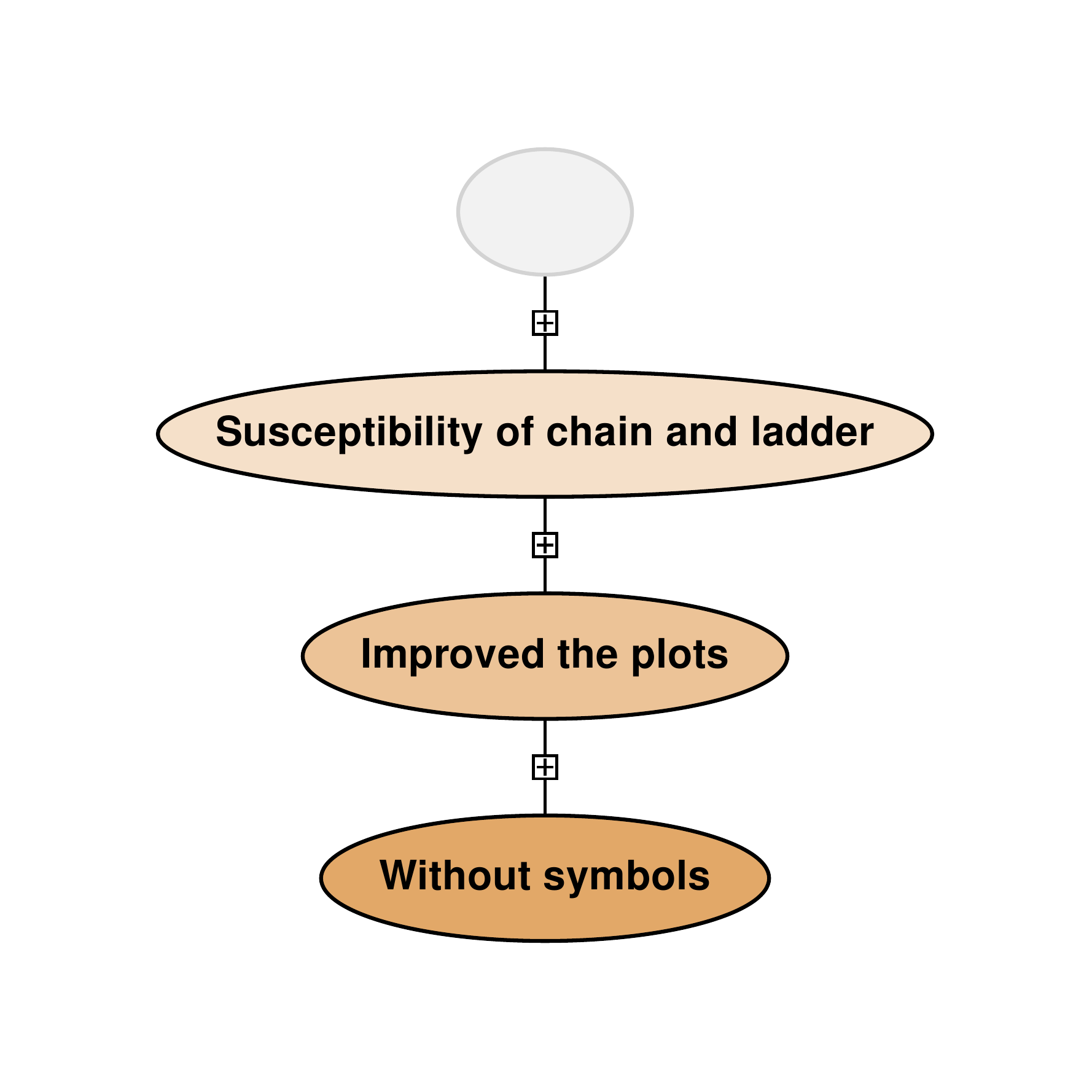}}
\caption{The version history tree of the workflow that created Fig.~\ref{fig:figure}. Each ellipsis corresponds to a version of the workflow, as shown in Fig.~\ref{fig:workflow}. Clicking the figure retrieves the vistrail including all workflow versions.}
\label{fig:history}
\end{center}
\end{figure}

\subsection{Caching and persistent storage}


One important piece of infrastructure provided by VisTrails is the
management of data both used and produced by computational processes,
including intermediate results. While parameters and workflow
structure may change from execution to execution, many of the
intermediate results can be reused without recomputation. Note that
with traditional scripts, users must manually record and recall
intermediate results, meaning it is often necessary to repeat the
entire computation. VisTrails supports efficient exploration of such
analyses by automatically caching intermediate results. The results
in the cache are identified by the signature of the upstream
subworkflow (a hash computed from all of the computations that led to
the value). If the signature exists in the cache, we need not compute
any of the upstream subworkflow but instead return the value in the
cache. If anything in the subworkflow changes, the signature will
also change, and VisTrails will recompute the value. VisTrails also
allows modules that are not deterministic to be flagged as
non-cacheable; note that all modules downstream of such modules must
also be recomputed each time.

In addition, VisTrails provides a persistence package to help users
manage their data and provide persistent caching. This package
provides storage and versioning of input, intermediate, and output
data to ensure that workflows are able to access their data. Data is
identified by unique IDs rather than file paths, and users can name
and annotate the data in order to facilitate searching. The data can
be linked to a local disk but is stored in a managed repository to
ensure that it is retained. VisTrails will create new versions if
data changes, and provides persistent caching using the same signature
scheme as in-memory caching. In contrast to the in-memory caching
mechanism, a persistent intermediate result can be recalled in later
sessions. Using the metadata stored in persistent store and
provenance information, it is possible to connect data with the
workflows that either used or generated it and vice
versa~\cite{koop@ssdbm2010}. In addition, users can not only trace
data lineage but also retrieve the exact input data---even if that
data was changed or removed on the filesystem.

\subsection{Sharing content and reproducible research}

One of the goals of the VisTrails system is to
facilitate the publication of scientific results. VisTrails and its social workflow repository
CrowdLabs greatly simplify the process of packaging workflows and
results for publication. By supporting the publication on wikis, 
web sites, or scientific documents (through LaTeX extensions such as in this paper), this framework allows users to create
documents whose digital artifacts (\eg figures) include a deep
caption: detailed provenance information which contains the
specification of the computational process (or workflow) and
associated parameters used to produce the
artifact. Figures~\ref{fig:figure} and \ref{fig:workflow} are examples
of this mechanism in action: both images were included by using
VisTrails directly and clicking on the images will download the
workflow with the parameters used to generate them. This workflow can then be used by the reader to reproduce the figures.

CrowdLabs is a social online workflow repository for the VisTrails
system. It combines workflow management tools with a scalable
infrastructure to allow scientists to collaboratively analyze and
visualize data. CrowdLabs is capable of hosting, executing, and
serving results from VisTrails workflows, including those that make
use of the ALPS VisTrails package. CrowdLabs also leverages provenance
information (\eg workflow/pipeline specifications, libraries,
packages, users, datasets and results) to provide a richer sharing
experience: users can search and query this information.

VisMashups~\cite{Santos09}, another component developed on top of VisTrails, is a system for simplifying the creation,
maintenance and use of customized workflows. It is also
integrated into CrowdLabs and enables users to modify parameters and
re-execute workflows from within their web browsers. This provides an
entry point for non-experts to explore and interact with complicated
simulation workflows. We will develop ALPS tutorials also as VisMashups, enabling the interactive exploration of tutorial problems with the reader's web browser, without the need to install ALPS.

\section{Python examples}

\begin{figure}
\begin{multicols}{2}
\begin{lstlisting}
# Lists:
tvalues = [0.05, 0.1, 0.2, 0.3, 0.4]
print tvalues[0]

# A list with different types
tvalues = [0.5, 'abc']
\end{lstlisting}
\columnbreak
\begin{lstlisting}
# Creating and accessing a dictionary
parms =
  { 
   'LATTICE' : 'chain lattice',
   'MODEL' : 'spin',
   'local_S' : 0.5,
   'T' : 0.01,
   'J' : 1 ,
   'THERMALIZATION' : 5000,
   'SWEEPS' : 50000,
   'L' : 60,
   'ALGORITHM' : 'loop'
  }

print parms['J']

\end{lstlisting}
\end{multicols}
\caption{Examples for creating lists and dictionaries in Python.}
\label{fig:listdict}
\end{figure}

The Python language and its use for the evaluation tools of ALPS are explained in Section~\ref{sct:python}. In this appendix, we give some examples for the usage of Python and the pyalps package.

Python contains a large number of built-in high-level data types, such as strings, integers, floating point numbers, lists, and dictionaries. The dictionary type in Python is called a {\tt dict}, and is a mutable group of key and value pairs. All the built-in container types in Python can store elements of different type at the same time. The {\tt dict} type is used in ALPS 2.0 to pass metadata about a simulation to the simulation execution commands. An example of this is is shown in Fig.~\ref{fig:python}, where the parameters for a series of Monte Carlo simulations are stored in the dict {\tt parms} and passed to the {\tt writeInputFiles} function of the pyalps package to write input files for the loop application. Fig.~\ref{fig:listdict} shows examples of how to use lists and dictionaries in Python.

The {\tt def} statement defines a function or method. An example of a simple Python function from the pyalps package is given in Fig.~\ref{fig:Pythonfunction}. This example also demonstrates the unusual feature of Python that tab indentation is used for block delimiters.

\begin{figure}
\begin{lstlisting}
def subtract_spectrum(s1,s2,tolerance=1e-12):
    res = pyalps.DataSet()
    res.props = s1.props

    for i in range(len(s1.x)):
        remove = False
        for j in range(len(s2.x)):
            if abs(s1.x[i]-s2.x[j]) < tolerance and abs(s1.y[i]-s2.y[j]) < tolerance:
                remove = True
                break
        if not remove:
            res.x = np.append(res.x,s1.x[i])
            res.y = np.append(res.y,s1.y[i])
    
    return res
\end{lstlisting}
\caption{Example of a Python function from the pyalps package. The function takes two {\tt DataSet} instances as input and returns a new {\tt DataSet} that contains all (x,y) values that occur in the first but not in the second. The intended use is for plotting of spectra, where states that occur in several quantum number sectors should sometimes be shown only once.}
\label{fig:Pythonfunction}
\end{figure}

In addition to the procedural programming paradigm, Python supports object-oriented programming. Since the {\tt DataSet} class is a crucial part of the pyalps package, a slightly reduced version is shown as example class in Fig.~\ref{fig:Pythonclass}. The declaration of a class is started by the {\tt def} statement. A constructor can be provided as {\tt \_\_init\_\_}; in general, method names that start and end with {\tt \_\_} define special methods, such as operators. As an example, the {\tt \_\_repr\_\_} method of the {\tt DataSet} is used for converting to a string for the {\tt print} statement.

\begin{figure}
\begin{lstlisting}
class DataSet(ResultProperties):
  def __init__(self):
    ResultProperties.__init__(self)
    
    self.x = np.array([])
    self.y = np.array([])
    
  def __repr__(self):
    return "x=%s\ny=%s\nprops=%s" % (self.x, self.y, self.props)
\end{lstlisting}
\caption{Simple example of a Python class from the pyalps package.}
\label{fig:Pythonclass}
\end{figure}
All non-static methods of a Python class must take the instance of the class as a first argument; this is commonly called {\tt self}.

\begin{figure}
\begin{lstlisting}
import pyalps
from pyalps.plot import plot
import  numpy as np
import matplotlib.pyplot as plt

ds = pyalps.DataSet()
ds.x = np.linspace(0,1,10)
ds.y = ds.x**2
ds.props['xlabel'] = '$\gamma$'
ds.props['ylabel'] = '$\gamma^2$'
ds.props['label'] = 'Example legend'

plot([ds])
plt.legend()
plt.show()
\end{lstlisting}
\caption{Plotting a simple function with pyalps.}
\label{fig:plotting}
\end{figure}

In Fig.~\ref{fig:plotting}, an example script is shown which creates a {\tt DataSet} instance, creates some data ($y = x^2, x \in [0,1]$) and makes a plot using plotting functionality from the pyalps package. We store the legend string and the axis labels for the final plot into the {\tt props} dict of the {\tt DataSet} instance. Note that the plotting function from pyalps is passed a list with one {\tt DataSet} -- this function, like most other pyalps functions, operates directly on lists of {\tt DataSet}s in order to assist the user in dealing with larger sets of data.

\section{ALPS Installation with CMake}
\label{sec:cmake}

\subsection{Prerequisites}

The following tools and libraries are required to build ALPS:

\begin{itemize}
\item The CMake build system~\cite{cmake} of version 2.8.0 or newer.
\item The BLAS~\cite{blasnetlib} and LAPACK libraries~\cite{lapack}. Ideally optimized versions for the target architecture should be used instead of the generic netlib versions.
\item The Boost C++ libraries version 1.43.0 or later (Boost 1.45.0 is included in one version of the source tarball)~\cite{boost}. Boost does not have to be installed; instead, the ALPS installation process can build the required Boost libraries if necessary.
\item The HDF5 library version 1.8~\cite{hdf5}.
\item A standard complying C++ compiler. ALPS has been tested with gcc versions 4.2, 4.3 and 4.4 as well as MSVC 9, IBM xlC++ 11.1, and Intel's icpc 10 and 11.
\end{itemize}
To build optional parts of ALPS it is recommended to install in addition:
\begin{itemize}
\item Python version 2.5 or 2.6~\cite{python}.
\item The  NumPy~\cite{numpy}, SciPy~\cite{scipy}, and matplotlib~\cite{matplotlib} Python packages.
\item The VisTrails scientific workflow and provenance management system~\cite{vistrails} and all its dependencies.
\item A Fortran 90 compiler for the {\tt tebd} code.
\end{itemize}

\begin{table}
\centering
\begin{tabular}{l|p{10cm}}
\hline 
{\it Name} &{\it Description} \\ \hline
{\tt Boost\_LIBRARY\_PATH} &Path to the Boost source tree. If Boost is not available as binary, this has to be set by the user when first running {\tt cmake}. \\ \hline
{\tt CMAKE\_C\_COMPILER} &{\tt C} compiler. This is determined from the system path or environment variable {\tt CC}. \\ \hline
{\tt CMAKE\_CXX\_COMPILER} &{\tt C++} compiler. This is determined from the system path or environment variable {\tt CXX}. \\ \hline
{\tt PYTHON\_INTERPRETER} &Path of the Python interpreter that {\tt pyalps} should be built for. If not set, a default be determined or Python is disabled. \\ \hline
\end{tabular}
\caption{List of CMake variables that have to be provided when first running {\tt cmake}. These cannot be changed afterwards. \label{table:cmake_variables}}
\end{table}

\begin{table}
\centering
\begin{tabular}{l|p{10cm}}
\hline
{\it Name} &{\it Description} \\ \hline
{\tt CMAKE\_INSTALL\_PREFIX} &Installation prefix. Default is {\tt /opt/alps}. \\ \hline
{\tt VISTRAILS\_APP\_DIR} &Path to the directory where VisTrails is installed. \\ \hline
{\tt ALPS\_BUILD\_APPLICATIONS} &Enable or disable build process for ALPS applications. \\ \hline
{\tt ALPS\_BUILD\_EXAMPLES} &Enable or disable build process for ALPS examples. \\ \hline
{\tt ALPS\_BUILD\_TESTS} &Enable or disable build process for ALPS tests. \\ \hline
{\tt ALPS\_BUILD\_PYTHON} &Enable or disable build process for {\tt pyalps}. \\ \hline
{\tt HDF5\_LIBRARIES} &Path of the HDF5 libraries. \\ \hline
{\tt HDF5\_INCLUDE\_DIR} &Path to HDF5 header files. \\ \hline
{\tt BLAS\_LIBRARY} &Path of the BLAS library. \\ \hline
{\tt LAPACK\_LIBRARY} &Path of the LAPACK library. \\ \hline
\end{tabular}
\caption{List of CMake variables that can be changed at any time. Some of these are mandatory, i.e., if ALPS is unable to determine a value automatically, an error will be generated. \label{table:cmake_variables2}}
\end{table}

\subsection{The CMake build system}

While we generally recommend the use of the binary installation packages whenever available, the source installation using the CMake build system may offer several advantages to experienced users:
\begin{itemize}
\item Specific compiler versions and optimization flags can be provided.
\item The binary installers are tied to a specific version of Python, whereas the source installation can be built against the version of Python that the user prefers.
\item Custom versions of external libraries such as the BLAS and LAPACK libraries can be provided.
\end{itemize}

The CMake build process is controlled by a set of variables describing the build environment, paths, and targets (libraries and programs) to be built. Most of these variables are automatically determined in the top-level file {\tt CMakeLists.txt}. Each source directory also contains such a file with prescriptions for building certain targets given the global configuration variables. A major advantage of CMake over other build systems such as the {\tt autotools} is that it allows the user easy access and modification of all variables via a graphical interface, the tool {\tt ccmake} or by directly editing the file {\tt CMakeCache.txt}.

With a few exceptions (which are listed below), all CMake variables can be changed at any point during the installation. CMake will be able to determine the effects of changing a variable and, if necessary, repeat the compilation of parts of ALPS given the new configuration. This gives the user the possibility to change aspects of the configuration without repeating the entire installation process.

Some of the most important CMake variables for the ALPS build system are listed in Tables~\ref{table:cmake_variables} and \ref{table:cmake_variables2}. Note that library paths have to be given as absolute paths; if several libraries are required, e.g. for BLAS, the paths should be separated with semicolons. An important point is linking against the correct versions of BLAS and LAPACK, since this may seriously affect the performance of some applications. ALPS will try to determine automatically which libraries should be used and will give optimized libraries like Intel's MKL precedence over standard libraries. Nevertheless, it is recommended to check whether this has been recognized correctly.

\subsection{Installation steps}

It is strongly recommended to perform the configuration and compilation of ALPS in a separate directory ({\it build directory}). Running any CMake commands in the source directory will likely cause problems in later stages.
For concreteness, we will assume in the following that the ALPS source can be found in {\tt /home/user/alps} and Boost sources can be found in {\tt /home/user/boost\_1\_43\_0}. For other libraries, we will assume that they are installed in standard system paths. All commands should be executed in the build directory, which initially must be empty.

The first step is to create an initial CMake configuration:
{\small
\begin{verbatim}
cmake -D Boost_ROOT_DIR:PATH=/home/user/boost_1_43_0 /home/user/alps
\end{verbatim} }
Additional variables can be passed with the {\tt -D} flag. For example, if the HDF5 libraries are not found in a standard path, Python should not be used, and a non-standard compiler should be used, the following commands need to be executed:
{\small
\begin{verbatim}
export CC=/opt/local/bin/gcc-mp-4.4
export CXX=/opt/local/bin/g++-mp-4.4
cmake -D Boost_ROOT_DIR:PATH=/home/user/boost_1_43_0 \
-D HDF5_LIBRARIES=/home/user/HDF5/lib/libhdf5.so \
-D HDF5_INCLUDE_DIR=/home/user/HDF5/include \
-D ALPS_BUILD_PYTHON=OFF /home/user/alps
\end{verbatim} }
Please note that {\tt cmake} should always be run from an empty directory. If an error occurs, please clear all files from the build directory before running {\tt cmake} again. By default, CMake will generate a set of UNIX Makefiles. It is also possible to use other generators; for example, project files for XCode or KDevelop can be created.

In the next step, additional variables can be customized by either editing the file {\tt CMakeCache.txt} directly, running the command {\tt ccmake} or invoking the CMake GUI.

Now the standard procedure is to build, test, and install ALPS using
{\small \begin{verbatim}
make
make test
make install
\end{verbatim}}

The VisTrails packages can also be installed from the CMake installation process. On UNIX or Linux platforms, it is sufficient to set the CMake variable {\tt VISTRAILS\_APP\_DIR}. If, for example, the file {\tt vistrails.py} from the VisTrails installation is found in the directory {\tt /home/user/vistrails/vistrails/}, {\tt VISTRAILS\_APP\_DIR} should be set to {\tt /home/user/vistrails/}.
On Mac OS X, a more advanced build type is needed since VisTrails delivers a specific version of Python. For instructions for this special case, we refer to our website.

\section{The ALPS XML schemas}
\label{sec:xml}

\begin{figure}[tb]
\begin{lstlisting}
<PARAMETERS>
  <PARAMETER name="LATTICE"> square </PARAMETER>
  <PARAMETER name="MODEL">     spin </PARAMETER>
  <PARAMETER name="L">           10 </PARAMETER>
  <PARAMETER name="T">          0.5 </PARAMETER>
</PARAMETERS>
\end{lstlisting}
\caption{Excerpt from an XML file for simulation input parameters .}
\label{fig:parm}
\end{figure}
\begin{figure}[tb]
\begin{lstlisting}
<SCALAR_AVERAGE name="Susceptibility">
  <MEAN>                     421.3 </MEAN>
  <ERROR converged="yes">     1.54 </ERROR>
  <VARIANCE>              1.06e+05 </VARIANCE>
</SCALAR_AVERAGE>
\end{lstlisting}
\caption{Excerpt from an XML file for simulation results of the uniform susceptibility in a Monte Carlo simulation.}
\label{fig:result}
\end{figure}

XML is the format of choice for text-based ALPS files. Unlike simple formats, where the location of a number in the file specifies its meaning (e.g. the first number is system size, the second temperature), XML specifies the meaning of data through meta-information provided by markup with tags, as shown in figure \ref{fig:parm} for input parameters and in  figure \ref{fig:result} for simulation results. The meaning of these files is easy to decipher even years after the simulation, unlike many other formats. A resemblance to HTML is no chance, since HTML (XHTML) is indeed also an XML format. Instead of looking directly at the long (and sometimes ugly) output files in XML format, the XML files can be easily transformed to other formats using XSLT transformations\cite{xslt} and viewed, e.g. directly as HTML in a web browser, or printed as plain text. 

\begin{figure}
\begin{lstlisting}
<LATTICEGRAPH name = "square">
  <FINITELATTICE>
    <LATTICE dimension="2"/>  
    <EXTENT dimension="1" size="L"/>
    <EXTENT dimension="2" size="L"/>
    <BOUNDARY type="periodic"/>  
  </FINITELATTICE>
  <UNITCELL>
    <VERTEX/>
    <EDGE>
      <SOURCE vertex="1" offset="0 0"/>
      <TARGET vertex="1" offset="0 1"/>
    </EDGE>
    <EDGE>
      <SOURCE vertex="1" offset="0 0"/>
      <TARGET vertex="1" offset="1 0"/>
    </EDGE>
  </UNITCELL> 
</LATTICEGRAPH>
\end{lstlisting}
\caption{The definition of a square lattice with one site (vertex) per unit cell and bonds (edges) only to nearest neighbors. After describing the dimension, extent and boundary conditions of the Bravais lattice in the {\tt <FINITELATTICE>} element, the unit cell including the bonds in the lattice is defined.}
\label{fig:lattice}
\end{figure}
\begin{figure}
\begin{lstlisting}
<BASIS name="spin">
  <SITEBASIS>
    <QUANTUMNUMBER name="S" min="1/2" max="1/2"/>
    <QUANTUMNUMBER name="Sz" min="-S" max="S"/>
    <OPERATOR name="Splus" matrixelement="sqrt(S*(S+1)-Sz*(Sz+1))">   
      <CHANGE quantumnumber="Sz" change="1"/>
    </OPERATOR>
    <OPERATOR name="Sminus" matrixelement="sqrt(S*(S+1)-Sz*(Sz-1))">  
      <CHANGE quantumnumber="Sz" change="-1"/>
    </OPERATOR>
    <OPERATOR name="Sz" matrixelement="Sz"/>  
  </SITEBASIS>
</BASIS>

<HAMILTONIAN name="spin">
  <BASIS ref="spin"/>
  <SITETERM> -h*Sz </SITETERM>   
  <BONDTERM source="i" target="j">
    Jxy/2*(Splus(i)*Sminus(j) + Sminus(i)*Splus(j)) + Jz*Sz(i)*Sz(j)
  </BONDTERM>
</HAMILTONIAN>
\end{lstlisting}
\caption{The definition of a spin-1/2 $XXZ$ spin Hamiltonian with one type of exchange coupling only: $H=-h\sum_iS_i^z+\sum_{\langle i,j\rangle} \left((J_{\rm xy}/2)(S_i^+S_j^-+S_i^-S_j^+)+J_zS_i^zS_j^z\right)$.  After describing the local basis for each site and operators acting on it, the Hamiltonian is defined.}
\label{fig:model}
\end{figure}

Besides XML schemas for parameter input and output of results we have developed XML schemas for the description of lattices and models. Examples are shown in figures \ref{fig:lattice} and \ref{fig:model}. For detailed specification of the formats see our XML web page.\cite{xmlschema} 

Providing an input file (such as the one where an excerpt is shown in figure \ref{fig:parm} together with lattice and model definitions in figures  \ref{fig:lattice} and \ref{fig:model} to any of the ALPS applications will run that application on the given model (provided the application supports that type of models) and returns an output file containing data such as shown in figure \ref{fig:result}. 

\subsection{Lattice definitions}

\begin{figure}
\begin{lstlisting}
<LATTICEGRAPH name = "depleted inhomogeneous square lattice">
  <FINITELATTICE>
    <LATTICE dimension="2"/>  
    <EXTENT dimension="1" size="L"/>
    <EXTENT dimension="2" size="L"/>
    <BOUNDARY type="periodic"/>  
  </FINITELATTICE>
  <UNITCELL>
    <VERTEX/>
    <EDGE>
      <SOURCE vertex="1" offset="0 0"/>
      <TARGET vertex="1" offset="0 1"/>
    </EDGE>
    <EDGE>
      <SOURCE vertex="1" offset="0 0"/>
      <TARGET vertex="1" offset="1 0"/>
    </EDGE>
  </UNITCELL> 
  <INHOMOGENEOUS><VERTEX/></INHOMOGENEOUS>}
  <DEPLETION>
    <VERTEX probability="DEPLETION" seed="DEPLETION_SEED"/>
  </DEPLETION>
</LATTICEGRAPH>
\end{lstlisting}

\caption{The definition of a square lattice with one site (vertex) per
unit cell and bonds (edges) only to nearest neighbors. First the
dimension, extent, and boundary conditions of the Bravais lattice are
described in the {\tt <FINITELATTICE>} element, then the unit cell
including the bonds in the lattice is defined. The new feature shown here is the specification of an inhomogeneous lattice, allowing to specify different couplings on the sites (in this example), and the depletion of the vertices of the lattices.}
\label{fig:deplattice}
\end{figure}
Besides the ability to specify translation invariant lattices as shown in Fig. \ref{fig:lattice}, starting in release 1.3 ALPS has included the possibility to specify inhomogeneity and depletion. While in a regular lattice, the model Hamiltonian is the same for all bonds or sites with the same type, in an inhomogeneous lattice the couplings in the model can be different for each bond or site. Examples are disordered systems, with randomly chosen couplings such as spin glasses, or systems in spatially varying trapping potentials, such as optical lattices in harmonic traps.
Inhomogeneities can be specified for vertices (sites) and edges (bonds), either for all (such as for all vertices in figure \ref{fig:deplattice}), or only for one type of vertex or edge. 

One can also randomly deplete the lattice by removing a fraction of sites or bonds is randomly removed from a lattice. Currently only site depletion is implemented, but additional types of depletion can be added easily if  the need arises. Please contact the authors if you need that feature. In the example in figure  \ref{fig:deplattice}, the fraction of depleted sites is specified to be passed in the {\tt DEPLETION} input parameter, and the random number generator seed {\tt DEPLETION\_SEED} can be changed to give different random realizations of the depletion pattern.

For more examples of lattices we refer the reader to the file {\tt lattices.xml} included in ALPS that contains a number of commonly used lattices. A detailed discussion of the creation of lattice files is presented on the ALPS web page.

\subsection{Model definitions}

Figure \ref{fig:model} shows the simplest definition of an XXZ spin model with Hamiltonian
\begin{equation}
H=-h\sum_iS_i^z+\sum_{\langle i,j\rangle} \left((J_{\rm xy}/2)(S_i^+S_j^-+S_i^-S_j^+)+J_zS_i^zS_j^z\right)
\end{equation}

Figure \ref{fig:hubmodel} shows the definition of the Hamiltonian of a bosonic Hubbard model in a harmonic trap
 \begin{eqnarray}
 H&=&-\mu\sum_in_i+\frac{U}{2}\sum_in_i(n_i-1)   \label{eq:model} \\
    && -t \sum_{\langle
    i,j\rangle} \left(b_i^\dag b_j + b_j^\dag b_i\right) + \frac{K}{2}\sum_i(x_i^2+y_i^2). \nonumber
  \end{eqnarray}
One feature shown in this example, is the ability to specify composite site and bond operators, such as the {\tt double\_occupancy} site term or the {\tt boson\_hop} bond term. These site and bond operators can be used in the specification of the model and of measurements.
  The other new feature, also shown in the example, is the ability, in conjunction with the specification of an inhomogeneous lattice,  to use site-dependent couplings, such as the harmonic trapping potential $K$ which depends on the $x$ and $y$ coordinates of the site. This feature has been used for the simulation of cold bosonic gases in optical lattices \cite{Wessel2004}.

\begin{figure}
\begin{lstlisting}
<BASIS name="boson">
<SITEBASIS>
  <PARAMETER name="Nmax" default="infinity"/>
  <QUANTUMNUMBER name="N" min="0" max="Nmax"/>
  <OPERATOR name="bdag" matrixelement="sqrt(N+1)">
    <CHANGE quantumnumber="N" change="1"/>
  </OPERATOR>
  <OPERATOR name="b" matrixelement="sqrt(N)">
    <CHANGE quantumnumber="N" change="-1"/>
  </OPERATOR>
  <OPERATOR name="n" matrixelement="N"/>
</SITEBASIS>
</BASIS>

<SITEOPERATOR name="double_occupancy" site="x">
  n(x)*(n(x)-1)/2
</SITEOPERATOR>

<BONDOPERATOR name="boson_hop" source="x" target="y">
  bdag(x)*b(y)+bdag(y)*b(x)
</BONDOPERATOR>

<HAMILTONIAN name="trapped boson Hubbard">
  <PARAMETER name="mu" default="0"/>
  <PARAMETER name="t" default="1"/>
  <PARAMETER name="U" default="0"/>
  <PARAMETER name="K" default="0"/>
  <BASIS ref="boson"/>
  <SITETERM site="i">
    -mu*n(i) + U*double_occupancy(i) + K/2*((x-L/2)^2 + (y-L/2)^2)
  </SITETERM> 
  <BONDTERM source="i" target="j">
    -t*boson_hop(i,j)
  </BONDTERM>
</HAMILTONIAN>
\end{lstlisting}
\caption{The definition of a bosonic Hubbard Hamiltonian (\ref{eq:model}) in a harmonic trap.  After
  describing the local basis for each site and operators acting on it, one can also define composite site and bond operators, such as the double occupancy and bosonic hopping terms, which are then used in the definition of the Hamiltonian, instead of writing all terms explicitly as in the example in Fig. \ref{fig:model}. Another feature shown here is the use of the parameters $x$ and $y$ in the couplings to create spatially-dependent couplings when combined with an inhomogeneous lattice.}
\label{fig:hubmodel}
\end{figure}

For more examples of models we refer the reader to the file {\tt lattices.xml} included in ALPS that contains a number of commonly used models. A detailed discussion of the creation of model files is presented on the ALPS web page.

\section*{References}

\bibliographystyle{iopart-num}
\bibliography{refs}
\end{document}